\documentclass[twocolumn]{aastex62}
\usepackage{graphicx}
\usepackage{amsmath,amssymb,wasysym}

\usepackage{graphicx}
\usepackage{dcolumn}
\usepackage{bm}

\usepackage{xcolor}

\begin{document}

\title{Turbulence vs. fire hose instabilities: 3-D hybrid 
expanding box simulations}

\author{Petr Hellinger}
\affiliation{Astronomical Institute, CAS, Bocni II/1401, Prague CZ-14100, Czech Republic}
\affiliation{Institute of Atmospheric Physics, CAS, Bocni II/1401, Prague CZ-14100, Czech Republic}
\email{petr.hellinger@asu.cas.cz}

\author{Lorenzo Matteini}
\affiliation{LESIA, Observatoire de Paris, Universit\'e PSL, CNRS, Sorbonne Universit\'e, Univ. Paris Diderot, Sorbonne Paris Cit\'e, place Jules Janssen 5, 92195 Meudon, France}

\author{Simone Landi}
 \affiliation{Dipartimento di Fisica e Astronomia, Universit\`a degli Studi di Firenze, via G. Sansone 1, Sesto Fiorentino 50019, Italy}
\affiliation{INAF, Osservatorio Astrofisico di Arcetri, Largo E. Fermi 5, Firenze 50125, Italy}

\author{Luca Franci}
\affiliation{School of Physics and Astronomy, Queen Mary University of London, London E1 4NS, UK}

\author{Andrea Verdini}
 \affiliation{Dipartimento di Fisica e Astronomia, Universit\`a degli Studi di Firenze, via G. Sansone 1, Sesto Fiorentino 50019, Italy}

\author{Emanuele Papini}
 \affiliation{Dipartimento di Fisica e Astronomia, Universit\`a degli Studi di Firenze, via G. Sansone 1, Sesto Fiorentino 50019, Italy}

\date{\today}

\begin{abstract}
The relationship between a decaying plasma turbulence and proton fire hose instabilities in a slowly expanding plasma
is investigated using three-dimensional (3-D) hybrid expanding box simulations.
We impose an initial ambient magnetic
field along the radial direction, and we start with an isotropic spectrum of
large-scale, linearly-polarized, random-phase Alfv\'enic fluctuations with zero cross-helicity.
 A turbulent cascade rapidly develops and
leads to a weak proton heating that is not sufficient to overcome the expansion-driven perpendicular cooling.
The plasma system eventually drives the parallel and oblique fire hose instabilities
that generate quasi-monochromatic wave packets that reduce the proton temperature anisotropy.  
The fire hose wave activity has a low amplitude with wave vectors quasi-parallel/oblique with
respect to the ambient magnetic field outside of
the region dominated by the turbulent cascade and 
is discernible in one-dimensional power spectra taken only in the direction quasi-parallel/oblique
with respect to the ambient magnetic field; at quasi-perpendicular angles
the wave activity is hidden by the turbulent background. 
These waves are partly reabsorbed by protons
and partly couple to and participate in the turbulent cascade. Their presence
reduces kurtosis, a measure of intermittency, and the  Shannon entropy
but increases the Jensen-Shannon complexity of magnetic fluctuations;
these changes are weak and anisotropic with respect 
to the ambient magnetic field and it's not clear if they can be used
to indirectly discern the presence of instability-driven waves. 
\end{abstract}

\section{Introduction}

The solar wind is a turbulent flow of weakly collisional plasma
and constitutes a natural laboratory for 
plasma turbulence \citep{brca13,alexal13}.
 Properties of plasma turbulence and its dynamics
remain an open challenging problem \citep{petral10,mave11}.
At large scales it
can be described by the magnetohydrodynamic (MHD) approximation,
 accounting for the dominant nonlinear coupling and
for the presence of the ambient magnetic field that introduces a preferred
direction \citep{shebal83,oughal94,verdal15}.
Around particle characteristic scales
 the plasma description has to be extended beyond MHD \cite[cf.,][and references therein]{papial19} and
a transfer of the cascading energy to particles is expected.
Observed radial profiles of proton temperatures indicate an important heating
which is often comparable to the estimated turbulent energy cascade rate \citep{macbal08,cranal09,hellal13}.
Alpha particles also need to be heated \citep{stanal19} whereas
for electrons it is much less clear because of their strong heat flux
\citep{stveal15}.

The nonlinearly coupled turbulent system is also characterized by non-Gaussian 
statistical properties, a phenomenon called intermittency. 
For a given quantity $Y$ (e.g., a component of the magnetic field, bulk velocity field)
and a separation length $l$, the increment, $\Delta Y$, the difference between
two points separated by $l$, has generally a non-Gaussian distribution. This non-Gaussianity
is well characterized by the excess kurtosis (flatness) that is close to zero on 
large scales (compatible with a Gaussian statistics) but it increases as $l$ 
decreases in the inertial ranges; behavior of the kurtosis in the sub-ion range
is not well known, some results indicate that it further increases \cite[e.g.,][]{alexal08,franal15b}
but some observations and numerical simulations indicate that the kurtosis
saturates and even decreases in the sub-ion range \cite[e.g.,][]{wual13,paraal18}.
The non-Gaussian statistics with extreme values is 
often associated with localized/coherent structures that 
appear to be important for the particle energization \citep{mattal15d};
these energization/dissipation processes could be responsible for the decrease of the kurtosis.

Recently, the Shannon entropy, $H$, and the Jensen-Shannon complexity, $C$ (see below
for definitions), two related 
quantities that measure the information content in time series
   and allows to discern chaotic versus stochastic behavior \citep{mamo13},
were used to characterize also turbulent fluctuations
observed in situ or in laboratory plasmas \citep{weckal15}. 
The time series of the solar wind magnetic field exhibit a relatively large 
normalized permutation (Shannon) entropy ($H\gtrsim0.9$), and a small
statistical (Jensen-Shannon)  complexity ($C\lesssim0.15$)  \citep{weckal15,weki19,olival19}.
In situ observations also indicate that
during the radial evolution the permutation entropy $H$ increases whereas
the complexity $C$ decreases
and these statistical parameters in the ($H,C$) plane  have properties similar to those
predicted for a fractional Brownian motion \citep{weki19}; interestingly, the fractional Brownian motion
may be regarded as a natural boundary between chaotic and stochastic processes \citep{mamo13};
however, the connection between these two statistical parameters and properties of the solar wind
turbulence is yet unclear.

The radial expansion further complicates the evolution of the solar wind turbulence.
 It induces an additional damping,
turbulent fluctuations decrease due to the expansion (as well as due to the turbulent decay)
slowing down the turbulent
cascade \cite[cf.,][]{grapal93,dongal14}. Furthermore, the expansion also introduces
another preferred direction (the radial one)  \citep{vegr15} and 
 the characteristic particle
scales change with the radial distance.
The understanding of the complex nonlinear properties of plasma turbulence on particle scales
is facilitated via a numerical approach \citep{serval15,franal18a}.
Direct kinetic simulations of turbulence show that particles are indeed on average heated by the cascade
\citep{paraal09,mava11,wual13,franal15b,arzaal19}, and, moreover, turbulence leads locally to
complex anisotropic and nongyrotropic distribution functions
\citep{valeal14,serval15}. On the other hand, the solar wind expansion
naturally generates particle temperature anisotropies \citep{mattal07,mattal12}. The anisotropic and nongyrotropic features
may be a source of free energy for kinetic instabilities.
There are indications that such instabilities are active
in the solar wind. Apparent bounds on ion parameters
are observed that are relatively in agreement with
theoretical kinetic linear predictions \citep{garyal01,hellal06,marual12,mattal13,kleinal18}.
Furthermore, wave activity driven by the kinetic instabilities associated with
the ion temperature anisotropies and/or differential velocity 
is often observed 
\citep{jianal09,garyal16,wickal16}. This activity is usually in the form
of a narrow-band enhancement of the magnetic power spectral density at ion scales
`on top of' the background turbulent magnetic field; these fluctuations
tend to be relatively coherent \citep{lional16}.

\cite{hellal15} used a two-dimensional (2-D) hybrid expanding box (HEB) model to
study the relationship between the proton oblique fire hose
instability and plasma turbulence. They showed that
the instability and plasma turbulence can coexist and
that the instability bounds the proton temperature
anisotropy in an inhomogeneous/nonuniform turbulent system. 
The results of \cite{hellal15} are, however,  
strongly constrained by the 2-D geometry with
the ambient magnetic field being perpendicular to the
simulation plane and one expects that unstable modes
appear at quasi-parallel/oblique angles with
respect to the background magnetic field
based on the linear (uniform and homogeneous) theory \cite[cf.,][]{gary93,klho15}.
 In this paper
we extend the work of \cite{hellal15} using a 3-D version of the HEB model.
We investigate the real-space and spectral properties
of turbulent fluctuations and the impact of fire hose
driven waves.
 We also analyze the effects of
this wave activity on the statistical properties (intermittency,
permutation entropy, complexity) of turbulence and test if these
quantities could be used as a mean to discern the presence
of the fire hose waves.

\section{Numerical Code}
\label{code}
In this paper we use  a 3-D version of the HEB model 
implemented in the numerical code CAMELIA
\citep{franal18b}
that allows us
to study self-consistently different physical processes at ion scales \cite[cf.,][]{ofma19}.
The expanding box uses coordinates co-moving with the solar
wind plasma, assuming 
a constant solar wind radial velocity $v_{sw}$, and
approximating the spherical coordinates by the Cartesian ones \citep{hetr05}.
The radial distance $R$ of the box 
follows
$R = R_0 (1+ t/t_{\text{exp}})$
where $R_0$ is the initial position and $t_{\text{exp}}=R_0/v_{sw}$
 is the (initial) characteristic expansion time.
Transverse scales of the simulation box (with respect to the radial direction) increase
$\propto R$ whereas the radial scale remains constant.
The model uses the hybrid approximation where electrons are
considered as a massless, charge neutralizing fluid and
ions are described by a particle-in-cell model \citep{matt94}.
Fields and moments are defined on a 3-D grid
$512 \times 512 \times 256$; periodic boundary conditions are assumed.
The initial spatial resolution is
$\Delta x=\Delta y=  0.25 d_{i}$, $\Delta z= 0.5 d_i$ where $d_{i}=v_{A}/\omega_{ci}$ is
the initial proton inertial length ($v_{A}$: the initial Alfv\'en velocity,
$\omega_{ci}$: the initial proton gyrofrequency).
There are $400$ macroparticles per cell
for protons which are advanced
with a time step $\Delta t=0.05/\omega_{ci}$
while the magnetic field
is advanced with a smaller time step $\Delta t_B = \Delta t/10$.
The initial ambient magnetic field is directed along the $z$ (radial) direction,
$\boldsymbol{B}_{0}=(0,0,B_{0})$, and
we impose a
continuous expansion in $x$ and $y$ directions with the initial expansion time
$t_{\text{exp}}=10^4 \omega_{ci}^{-1}$.
Due to the expansion with the strictly radial magnetic field
the ambient density and the magnitude of the ambient magnetic field
decrease as $\langle n\rangle  \propto \langle B\rangle \propto  R^{-2}$ (the proton
inertial length $d_{i}$ increases $\propto R$,
the ratio between the transverse sizes and $d_{i}$ remains constant
whereas  the ratio between the radial size and $d_{i}$ decreases as $R^{-1}$).

We set initially
the parallel proton beta  $\beta_{\|} =  2.4$
and the proton temperature anisotropy
 $T_{\perp}/T_{\|}=0.75$; for these parameters the plasma system
is stable with respect to the fire hose instabilities.
Electrons are assumed to be isotropic and isothermal with $\beta_{e} =   1$.
We initialize the simulation with an isotropic 3-D spectrum
of modes with random phases, linear Alfv\'en polarization ($\delta
\boldsymbol{B} \perp \boldsymbol{B}_0$), and vanishing correlation between
magnetic and velocity fluctuations (zero cross-helicity). These modes are in the range
 $0.02 \le k d_{i} \le 0.2$ and have a flat one-dimensional (1-D) 
(omnidirectional) power spectrum with rms fluctuations $=0.24 B_0$.
A small resistivity $\eta$ is used to avoid accumulation of
cascading energy at grid scales;
we set $\eta = 0.001 \mu_0 v_{A}^2/\omega_{ci}$
($\mu_0$ being the magnetic permittivity of vacuum).

\section{Simulation results}

\subsection{Global evolution}
\begin{figure}[htb]
\centerline{\includegraphics[width=8cm]{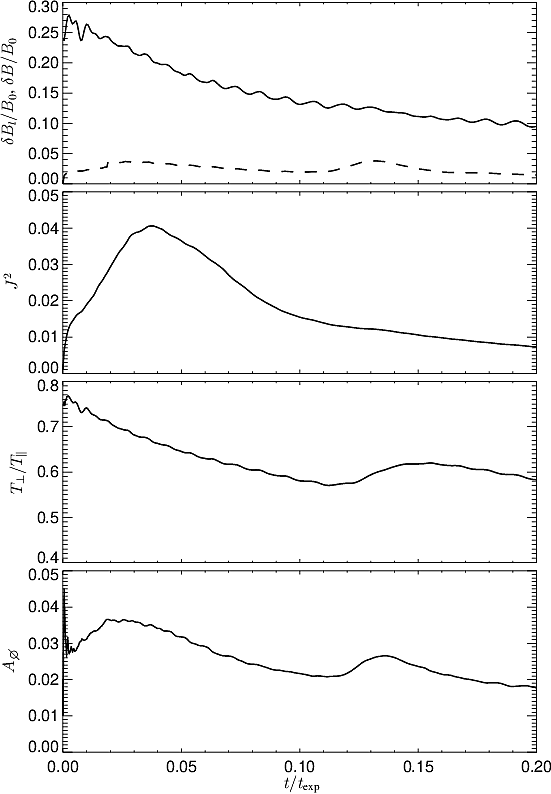}}
\caption{Evolution of 
 (from top to bottom) the total fluctuating magnetic field, $\delta B$ (solid line),
and the fluctuating magnetic field with large parallel wave vectors, 
$\delta B_l$ (dashed line), the 
mean square of the electric current $|\boldsymbol{J}|^2$,
the mean proton temperature anisotropy $T_\perp /T_\|$,
and the mean proton agyrotropy $A_{\diameter}$
 as functions of time. 
\label{hist}}
\end{figure}

The evolution of the system is shown in Fig.~\ref{hist} that
displays quantities averated over the simulation box:
 the (rms) total fluctuating magnetic field $\delta B$ and the fluctuating magnetic field
$\delta B_l$ with large parallel wave vectors (i.e., for
for $|k_\| | d_i>0.25$), 
the mean square of the electric current, $|\boldsymbol{J}|^2$, 
the mean proton temperature anisotropy $T_\perp /T_\|$,
and the mean proton agyrotropy $A_{\diameter}$
 as functions of time.
The agyrotropy $A_{\diameter}$ is defined here following \cite{scda08} 
as 
\begin{equation}
A_{\diameter}=2 \frac{|P_{\perp 1} - P_{\perp 2}|}{P_{\perp 1} + P_{\perp 2}}
\label{agyr}
\end{equation}
where $P_{\perp 1}$ and $P_{\perp 2}$ are the 
are the two eigenvalue components  of the (proton) pressure tensor
perpendicular to the local magnetic field.

The level of total magnetic fluctuations, $\delta B$ initially shortly
increases as the turbulent cascade develops and a part of the
kinetic proton energy is transformed to the magnetic one \cite[cf.,][]{franal15b}.
After that short transient period, $\delta B$
overall decreases and oscillates with
a small amplitude due to superposition of large-scale propagating
Alfv\'en modes. 
The amplitude of magnetic fluctuations with $|k_\| | d_i>0.25$, $\delta B_l$,
(top panel of Fig.~\ref{hist}, dashed line) initially increases,
reaches a local maximum around $t=0.025 t_{\text{exp}}$ as the turbulent cascade
develops. After that it decreases until 
it reaches another local maximum
 at about $t=0.13 t_{\text{exp}}$.
The mean amplitude of the current, $|\boldsymbol{J}|$, initially increases and reaches a
maximum at about $t=0.04 t_{\text{exp}}$ which indicates the presence
of a well-developed turbulent cascade \citep{mipo09,valeal14}.
After that, $|\boldsymbol{J}|$ decreases with a hint of a slower decrease
rate around $t=0.13 t_{\text{exp}}$.
The temperature
anisotropy, $T_\perp /T_\|$, after an initial transition
decreases (with weak oscillations) but it increases
from $t \sim 0.12 t_{\text{exp}}$ to $\sim 0.15 t_{\text{exp}}$ (and the system
becomes less anisotropic),
when modes with large parallel wave vectors appear;
after $t \sim 0.16 t_{\text{exp}}$, $T_\perp /T_\|$ decreases
again.
The agyrotropy, $A_{\diameter}$, starts from the initial
noise level $0.01$, strongly varies during the
short relaxation of the initial conditions and
then increases, reaches a maximum at around
$t=0.02t_{\text{exp}}$ and decreases. Later on, at the same time
as the proton anisotropy is reduced, the agyrotropy
increases again, reaches a maximum around $0.14t_{\text{exp}}$
then and decreases. The turbulent cascade creates velocity shears
at ion scales that naturally generate anisotropy as well as agyrotropy \citep{depe18}. 
Figure~\ref{hist} suggests that the reduction
of the temperature anisotropy occurring at $t \sim 0.12 t_{\text{exp}}$ -- $0.15 t_{\text{exp}}$
is related to the development of fire hose-like instabilities
generating fluctuations with large parallel wave vectors. While 
the fire hose instabilities
driven fluctuations reduce their source of free energy, they 
enhance the agyrotropy. 

\begin{figure}[htb]
\centerline{\includegraphics[width=8cm]{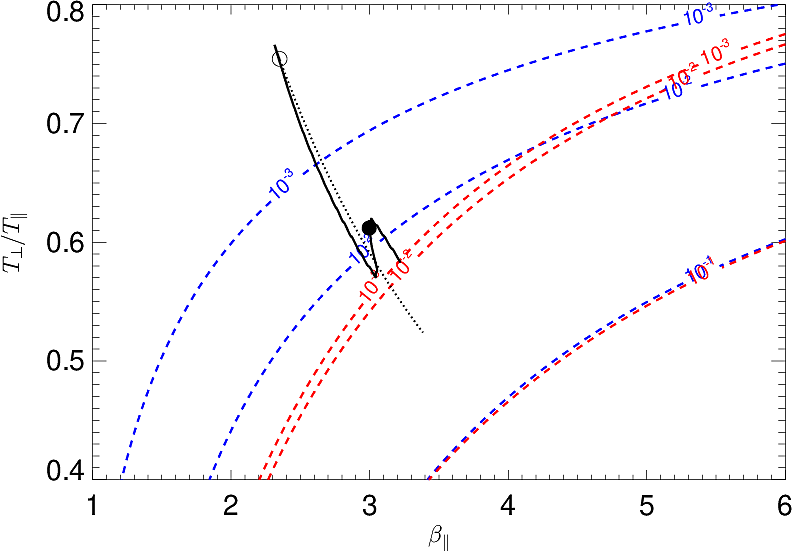}}
\caption{Evolution of the system in the $(\beta_\|,T_\perp /T_\|)$ space (solid line).
The empty circle denotes the initial condition whereas the full circle
denotes the time $t=0.14 t_{\text{exp}}$; the dotted line shows
the double adiabatic prediction for a corresponding system without turbulent
fluctuations. Blue and red dashed contours show the maximum growth rate
(normalized to $\omega_{ci}$)
of the parallel and  oblique fire hose instability (for a uniform and homogeneous bi-Maxwellian system), respectively. 
\label{anbe}}
\end{figure}

Figure~\ref{anbe} shows the evolution of the system in the $(\beta_\|,T_\perp /T_\|)$ space
that is useful to parametrize the linear stability of a (uniform and homogeneous) plasma
with respect to temperature anisotropy-driven instabilities  \citep{garyal01}.
Figure~\ref{anbe} shows that after the initial transient evolution,
$T_\perp /T_\|$ decreases and $\beta_\|$ increases somewhat similarly
to the double-adiabatic prediction ($T_\perp \propto R^{-2}$, $T_\| =\mathrm{const}$.).
During this phase, $\beta_\|$ increases whereas $T_\perp /T_\|$ decreases.
Eventually, the system reaches parameters that are unstable  (with respect to the parallel fire hose) in the corresponding 
uniform and homogeneous plasma. 
Before reaching the region unstable with respect to the oblique fire
hose instability the system moves towards the stable region and
$T_\perp /T_\|$ increases whereas $\beta_\|$ decreases. This reduction
of the proton anisotropy stops after some time and $T_\perp /T_\|$ decreases and $\beta_\|$ increases
until the end of the simulation.
The last part of the path of the system in the $(\beta_\|,T_\perp /T_\|)$ space
is very similar to results from HEB simulations without turbulent fluctuations
\cite[cf.,][]{hell17} when the system is under the influence of
the parallel and oblique fire hose instabilities. The latter is particularly
efficient in reducing the proton temperature anisotropy due
to its peculiar nonlinear evolution: the oblique fire hose generates
non propagating modes that only exist for a sufficiently strong
proton parallel temperature anisotropy. As these modes grow they scatter
protons and reduce their temperature anisotropy. Consequently,
they destroy the non propagating branch and
 transform (via the linear mode conversion) to propagating modes that are damped and
during this process they further reduce the temperature anisotropy \citep{hema00,hema01}.  
We observe a similar evolution in the present simulation.

\subsection{Physical space}

\begin{figure}[htb]
\centerline{\includegraphics[width=8cm]{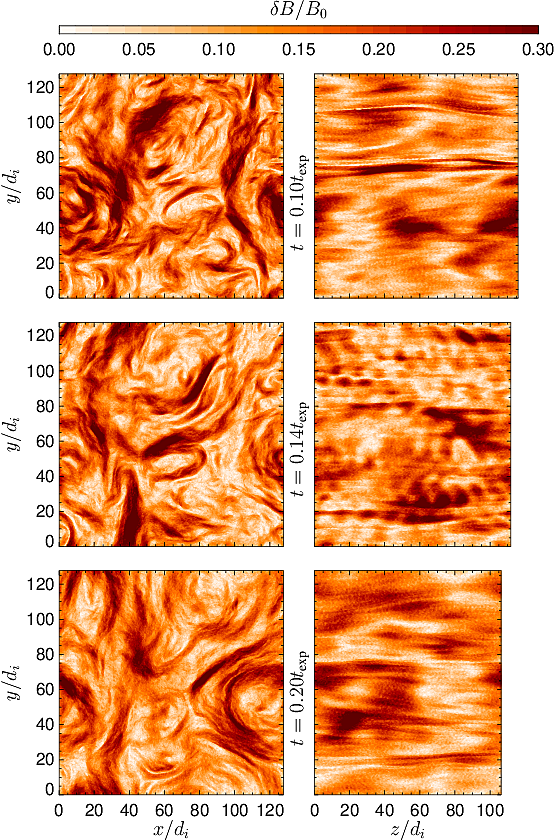}}
\caption{Color scale plots of 2-D cuts of $\delta B$ (normalized to $B_0$)
 as functions of $(x,y)$ (left) and $(z,y)$ (right) for $t=0.10 t_{\text{exp}}$ (top),
$t=0.14 t_{\text{exp}}$ (middle),  and $t=0.20 t_{\text{exp}}$ (bottom).
\label{bb}}
\end{figure}

We will now look at the properties of the simulated system at
three different times, at $t=0.10 t_{\text{exp}}$, before the onset
of the fire hose instabilities, at $t=0.14 t_{\text{exp}}$ around the
maximum activity of the instabilities, and at the end,
$t=0.20 t_{\text{exp}}$.
Figure~\ref{bb} shows the real space structure
of the magnetic fluctuations at the three times. It shows 2-D cuts of $\delta B$ 
 as functions of $(x,y)$  and $(z,y)$.
At $t=0.10 t_{\text{exp}}$ Figure~\ref{bb} (top panels) show a developed turbulence with
complex anisotropic properties. In the $(x,y)$ cut there are
signatures of nonlinear structures (such as current sheets and magnetic
islands) whereas the $(z,y)$ cut exhibits also large scale wavy
behavior \cite[cf.,][]{franal18a}.
At  $t=0.14 t_{\text{exp}}$ (Figure~\ref{bb}, middle panels)
the $(x,y)$ cut remains qualitatively the same as at  $t=0.10 t_{\text{exp}}$
but in the $(z,y)$ cut there is a clear wave activity, in the form of
localized quasi-coherent wave packets (with wavelengths of the order of $10 d_i$), on top of the
turbulent background.
At the end, $t=0.20 t_{\text{exp}}$, the $(x,y)$ cut is still qualitatively
unchanged and the $(z,y)$ cut is similar to that at $t=0.10 t_{\text{exp}}$. 
Note that a similar real space structure is also observed for other quantities
(e.g., ion bulk velocity, density, currents, temperature anisotropy/agyrotropy);
all these quantities are more uniform along the ambient magnetic field
or rather along the magnetic field lines.

\begin{figure}[htb]
\centerline{\includegraphics[width=8cm]{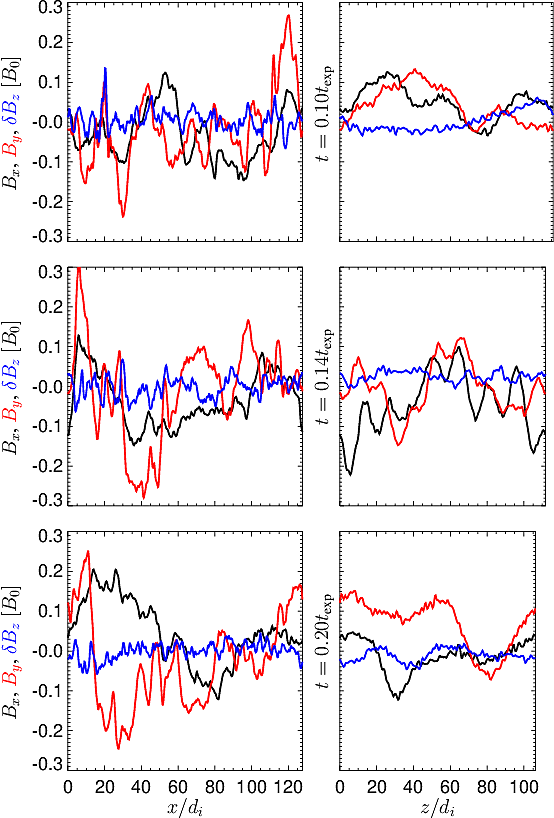}}
\caption{1-D cuts of the fluctuating magnetic field components (black) $B_x$, (red) $B_y$, and (blue) $\delta B_z$ (normalized to $B_0$)
 as functions of $x$ (and $y=z=0$, left) and $z$ (and $x=y=0$, right) for $t=0.10 t_{\text{exp}}$ (top),
$t=0.14 t_{\text{exp}}$ (middle),  and $t=0.20 t_{\text{exp}}$ (bottom).
\label{b1d}}
\end{figure}

To approach in situ 1-D observations we investigate 1-D spatial cuts through the simulation box 
for the three times.
Figure~\ref{b1d} shows 1-D plots of the fluctuating magnetic field components $B_x$, $B_y$, and $\delta B_z$ normalized to $B_0$
 as functions of $x$ (and $y=z=0$, left) and $z$ (and $x=y=0$, right).
As in Figure~\ref{bb} the perpendicular cuts (Figure~\ref{b1d}, left panels) are strongly stuctured
but it is hard to discern any similarities between the three times (except perhaps qualitative ones). In the parallel cuts
there are mostly large scale fluctuations at $t=0.10 t_{\text{exp}}$ and $t=0.20 t_{\text{exp}}$.
At $t=0.14 t_{\text{exp}}$ there are, on top of the large-scale turbulent fluctuations, shorter-wavelength wave packets
owing to the fire hose instabilities.
Figure~\ref{b1d} also shows that the compressible component $\delta B_z$ is weak with respect to $B_x$ and $B_y$
for the turbulent as well as fire hose fluctuations. Note that the radial ($z$) size of the simulation box
normalized to the proton inertial length decreases with time. 

\subsection{Spectral properties}
\begin{figure}[htb]
\centerline{\includegraphics[width=8cm]{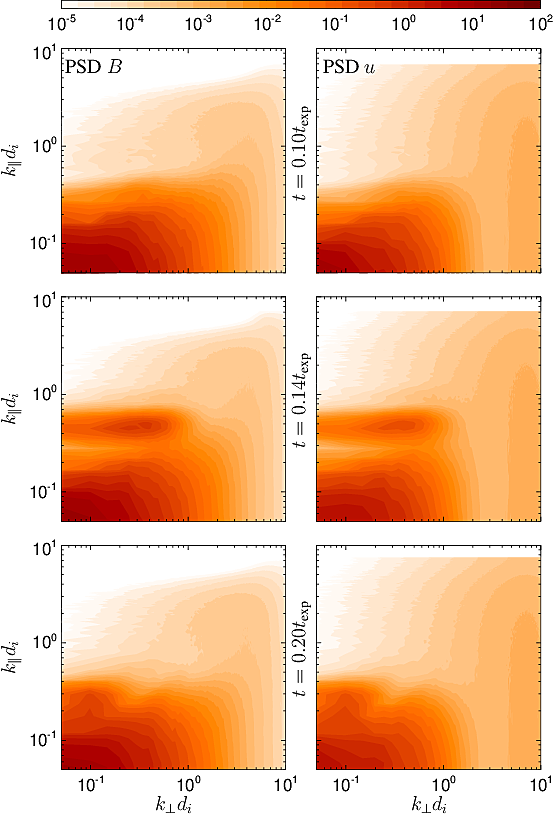}}
\caption{Color scale plots of the power spectral densities of the magnetic field  $\boldsymbol{B}$ (left) and 
the proton bulk velocity field $\boldsymbol{u}$
 (right)
 as functions of $k_\perp$ and $k_\|$ for $t=0.10 t_{\text{exp}}$ (top),
$t=0.14 t_{\text{exp}}$ (middle),  and $t=0.20 t_{\text{exp}}$ (bottom).
\label{psdbu}}
\end{figure}

Let us now investigate the spectral properties of the fluctuations shown
 in Fig.~\ref{bb}.
Figure~\ref{psdbu} shows the 2-D power spectral densities (PSDs) of
the magnetic field, $\boldsymbol{B}$, and the ion bulk velocity, $\boldsymbol{u}$,
 as functions of $k_\perp$ and $k_\|$ at the three times.
At $t=0.10 t_{\text{exp}}$ Figure~\ref{psdbu} (top panels)
demonstrates that the initial isotropic spectrum develops into a strongly
anisotropic one with a cascade preferably at strongly oblique angles
with respect to the ambient magnetic field \cite[cf.,][]{franal18a}.
At $t=0.14 t_{\text{exp}}$
(Figure~\ref{psdbu}, middle panels) there is a similar turbulent
spectrum and, moreover, a narrow band in $k_\|$ (at $k_\|\sim 0.5/d_i$)  enhancement at
quasi-parallel/oblique angles with respect to the background magnetic field.
At $t=0.20 t_{\text{exp}}$ this narrow band enhancement is not clearly discernible 
but there are some indications that some of the fluctuating energy
remained (around $k_\perp\sim 0.1 /d_i$ and $k_\|\sim 0.2 /d_i$).

\begin{figure}[htb]
\centerline{\includegraphics[width=8cm]{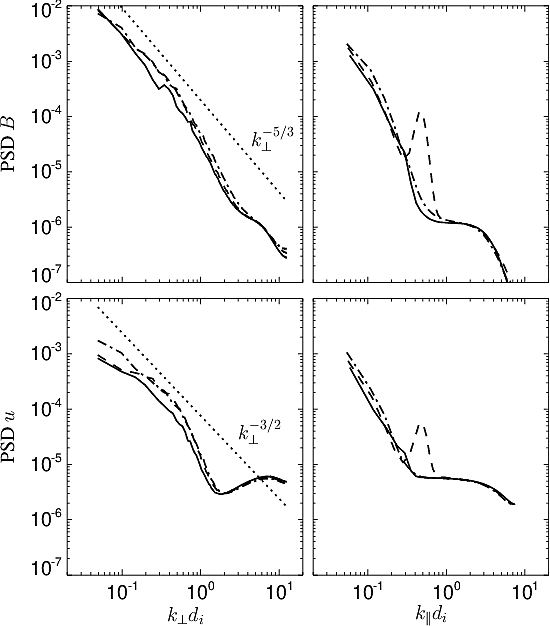}}
\caption{Reduced 1D power spectral densities of (top) $\boldsymbol{B}$  and (bottom)
$\boldsymbol{u}$
 as functions of (left) $k_\perp$ and (right) $k_\|$ for (dash-dotted) $t=0.10 t_{\text{exp}}$,
(dashed) $t=0.14 t_{\text{exp}}$,  and (solid) $t=0.20 t_{\text{exp}}$. The dotted lines 
show (top left) a Kolmogorov-like spectrum $\propto k_\perp^{-5/3}$
and (bottom left) a spectrum $\propto k_\perp^{-3/2}$. 
\label{psd1}}
\end{figure}

It is interesting to look at the reduced 1-D power spectra since
these correspond to what is usually obtained from observational
spacecraft data.
Figure~\ref{psd1} shows the comparison between 
 the reduced parallel and perpendicular 1D power spectral densities of 
$\boldsymbol{B}$  and $\boldsymbol{u}$
at the same three times as in Fig.~\ref{bb}. 
In the perpendicular direction, the PSDs of $\boldsymbol{B}$ and $\boldsymbol{u}$
 as functions
of $k_\perp$ (Figure~\ref{psd1}, left panels) 
show a little variation between the three times. The
fire hose wave activity is not discernible here. 
The PSDs of $B$ and $u$ exhibit a steepening at ion scales
and some noise at smaller scales (numerical
noise is especially visible in the velocity fluctuations due to the limited number of
particles per cell).
On the other hand, in the parallel direction, the PSDs of  $\boldsymbol{B}$ and $\boldsymbol{u}$ as functions
of $k_\|$ (Figure~\ref{psd1}, right panels) exhibit a clear narrow peak
at  $t=0.14 t_{\text{exp}}$ (dashed line) compared to both $t=0.10 t_{\text{exp}}$
and $t=0.20 t_{\text{exp}}$. Also here the small scale fluctuations are dominated by
numerical noise.
Further analysis shows that only for angles between
the ambient magnetic field and the wave vector below  $\sim 40^\mathrm{o}$
the fire hose fluctuations are discernible in 
 the corresponding reduced 1-D power spectra.

\begin{figure}[htb]
\centerline{\includegraphics[width=8cm]{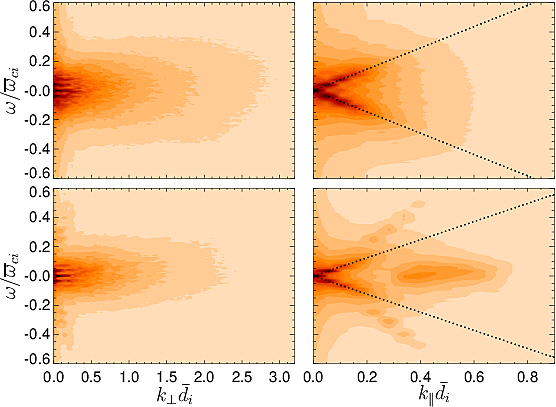}}
\caption{Spatio-temporal spectral properties of magnetic fluctuations: reduced
2-D spectra of $\boldsymbol{B}$ 
as functions of (left) $k_\perp$ and $\omega$ and (right) $k_\|$ and $\omega$ 
for (top) $t=0 \div 0.1 t_{\text{exp}}$ and (bottom) $t=0.1 \div 0.2 t_{\text{exp}}$.
The wavevectors are normalized to the mean (over the given time interval) inertial length $\bar{d}_i$ and
the frequency is normalized to the mean cyclotron frequency $\overline{\omega}_{ci}$.  
The dotted lines denote the dispersion relation of Alfv\'en waves.
\label{freq}}
\end{figure}

Figure~\ref{freq} shows the reduced parallel and perpendicular
2-D spatio-temporal spectra of magnetic fluctuations
during two phases, $t=0 \div 0.1 t_{\text{exp}}$ and $t=0.1 \div 0.2 t_{\text{exp}}$.
During the first, turbulent-only phase, the perpendicular
spectrum $\delta B(k_\perp,\omega)$ is broadly distributed
around the zero frequency; the spread is very large for
$k_\perp$ close to $0$.  
During the second, fire-hose phase the
perpendicular spectrum is similar but somewhat weaker.
The behaviour of the parallel spectrum $\delta B(k_\|,\omega)$ is different.
During the first phase the spectrum (for small $k_\|$) clearly exhibit Alfv\'enic dispersion,
$\omega=\pm k_\| \bar{v}_A$ (shown by the dotted lines) where $\bar{v}_A$ is
the mean (over the given time interval) Alfv\'en velocity including the effects of the proton temperature anisotropy.
During the second phase, these Alfv\'enic  
fluctuations are reduced and the parallel spectrum further contains
fast-magnetosonic dispersive modes ($\omega\propto \pm k_\|^2$)
for $0.2 \lesssim k_\| \lesssim 0.4$ and weakly or non propagating modes
at $0.3 \lesssim k_\| \lesssim 0.6$. The former modes are likely due
to the parallel fire hose 
\citep{garyal98,mattal06} whereas the latter 
are likely related to the oblique fire hose
\citep{hetr08}.

\subsection{Velocity distribution function}

\begin{figure}[htb]
\centerline{\includegraphics[width=8cm]{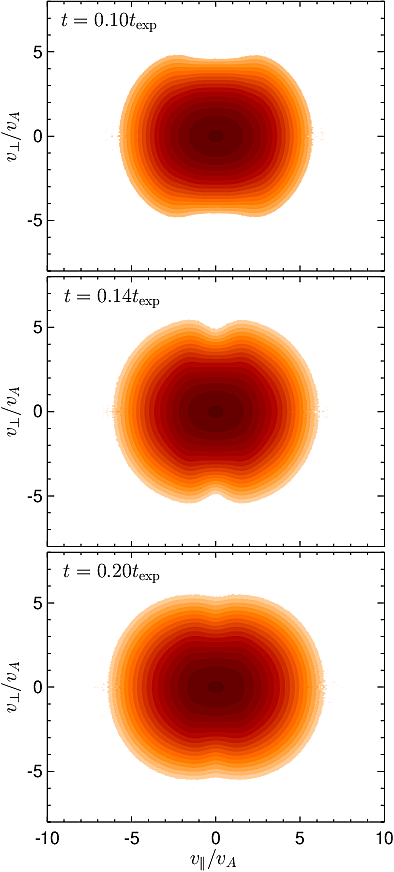}}
\caption{Gyro-averaged proton velocity distribution function $f$
as a function of parallel and perpendicular velocities $v_\|$ and $v_\perp$
(with respect to the background magnetic field) for (top) $t=0.10 t_{e}$,
(middle) $t=0.14 t_{e}$,  and (bottom) $t=0.20 t_{e}$.
\label{vdf}}
\end{figure}

The analysis above indicates that the expansion 
drives the system unstable with respect to the parallel and oblique
fire hose instabilities despite the presence of a developed anisotropic turbulent cascade.
The fire hose wave activity appears at quasi-parallel/oblique angles with
respect to the background magnetic field, outside the region in 
$(k_\perp,k_\|)$ where the turbulent fluctuations dominate.
The generated wave activity is likely partly damped (as expected
for the oblique fire hose) and partly blends into the turbulent 
fluctuations. 
Figure~\ref{vdf} shows the gyro-averaged proton velocity distribution function
at the different times. This figure shows that at $t=0.10 t_{e}$ is 
deformed with respect to the initial bi-Maxwellian distribution function 
due to the turbulent heating
\cite[cf.,][]{arzaal19}; at $v_\|\sim 3 v_A$ there are signatures
of the cyclotron heating owing to the  Alfv\'en (cyclotron) waves.
At $t=0.14 t_{e}$ the proton velocity distribution function
exhibits strong wings due to the standard and anomalous cyclotron 
resonance between the protons and fire hose waves. The signature
of this interaction is weakened at $t=0.20 t_{e}$ but still discernible.

\subsection{Statistical properties}

\begin{figure}[htb]
\centerline{\includegraphics[width=6cm]{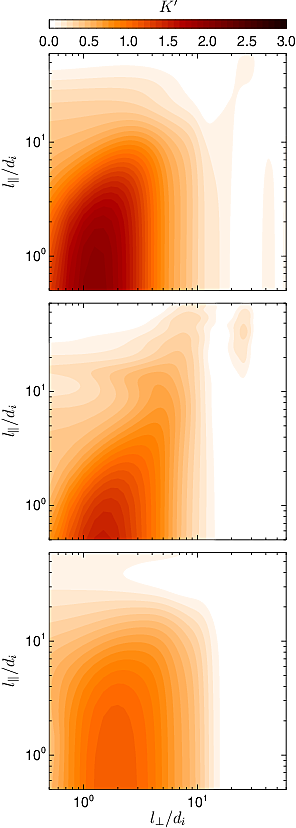}}
\caption{Color scale plots of the excess kurtosis $K^\prime$ of the increment  $\delta B_y$ 
 as functions of the separation $l_\perp $ and $l_\|$ for (top) $t=0.10 t_{\text{exp}}$,
(middle) $t=0.14 t_{\text{exp}}$,  and (bottom) $t=0.20 t_{\text{exp}}$.
\label{kurt}}
\end{figure}

It is interesting to look at how the fire hose wave activity affects
the statistical properties of turbulence. The turbulent
cascade leads to a non-Gaussian character of the fluctuations,
i.e., intermittency, which is likely affected by the wave activity generated by
the fire hose instabilities.
We calculated the excess kurtosis $K^\prime=K-3$ in
the separation space $(l_\perp,l_\|)$ 
over the simulation box 
for the increment $\delta B_y=B_y(x+l_\perp,y,z+l_\|)-B_y(x,y,z)$. 
The maximum value of $K^\prime$ appears at $l_\|=0$ and evolves with the time: initially 
reaches values of the order of 10 and after $t\simeq0.05 t_{\text{exp}}$
it stays about 3 and slowly decreases; there is no discernible effect of
the fire hose instabilities on it. $K^\prime$ is, however, very anisotropic.
Figure~\ref{kurt} shows $K^\prime$ as a function of $l_\perp$ and $l_\|$ at the three times $t=0.10 t_{\text{exp}}$
$t=0.14 t_{\text{exp}}$, and $t=0.20 t_{\text{exp}}$.
The kurtosis is similar at $t=0.10 t_{\text{exp}}$ and $t=0.20 t_{\text{exp}}$
(at the latter time is somewhat weaker): it's strongly dependent on $l_\perp$, increases as $l_\perp$ decreases but when approaching
the spatial resolutions $K^\prime$ saturates and decreases. 
At $t=0.14 t_{\text{exp}}$ the kurtosis is noticeably reduced at
the quasi-parallel direction, likely owing to the presence
of quasi-parallel/oblique waves.

Finally, it's interesting to look at how the fire hose wave activity
affects the permutation entropy $H$ and the Jensen-Shannon complexity $C$ 
\citep{lopeal95,bapo02,lambal04}.
The (normalized) permutation entropy is defined for
a time series where for $n$ consecutive points 
one calculates the probability for each permutation 
of larger/smaller ordering $p_j$. For this probability
vector $\boldsymbol{p}=\{p_j\}_{j=1,n!}$
the information/Shannon entropy is given as
\begin{equation}
S(\boldsymbol{p})= - \sum_{p_j\ne 0} p_j \ln p_j
\end{equation}
and the permutation entropy $H$ is normalized to its maximum value as
\begin{equation}
H(\boldsymbol{p})=\frac{S(\boldsymbol{p})}{\ln n!}.
\end{equation}
The Jensen-Shannon complexity is given by
\begin{equation}
C(\boldsymbol{p})=Q(\boldsymbol{p}) H(\boldsymbol{p})
\end{equation}
where
\begin{equation}
Q(\boldsymbol{p})=-\frac{2 S(\frac{\boldsymbol{p}+\boldsymbol{p}_e}{2})-S(\boldsymbol{p})-S(\boldsymbol{p}_e)}
{\frac{n!+1}{n!} \ln (n!+1)- 2\ln(2n!)+\ln n!}
\end{equation}
is a measure of the distance between $\boldsymbol{p}$
and
$\boldsymbol{p}_e$, the probability vector for equally probable permutations,
$\boldsymbol{p}_e=\{1/n!\}_{j=1,n!}$.

\begin{figure}[htb]
\centerline{\includegraphics[width=8cm]{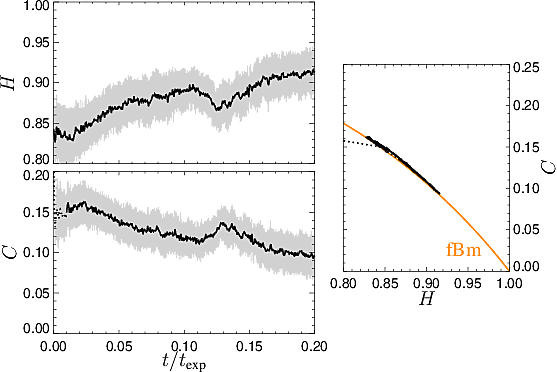}}
\caption{Evolution of
 (left top) the permutation entropy $H$ and  (left bottom) the
statistical complexity $C$ of $B_y$ component of the magnetic field
calculated at about $45^\mathrm{o}$ with respect to the ambient magnetic
field 
 as functions of time (mean values shown in black whereas the gray
area denote the mean values $\pm$ the standard deviation; dotted lines denote an initial
transition phase).
(right) Mean simulation data in the $(H,C)$ plane (black line)
compared to the prediction for the fractional Brownian motion
(orange line); dotted line denotes the initial transient phase.
\label{hc}}
\end{figure}

The two statistical quantities depend on many parameters (the scale, number of points, etc.)
and is also affected by the noise level
\citep{lambal04}. Having a non-negligible noise level and a limited box size we 
take $8\times8$ 1-D cuts (equidistantly separated) and calculate $H$ and $C$
for each cut every $5 \omega_ci$ using 4-point statistics ($n=4$ in the definitions above).
Figure~\ref{hc} shows the evolution of the mean permutation entropy $H$ and 
the mean statistical complexity $C$ (and their standard deviations)
 of the $B_y$ component of the magnetic field 
obtained for 1-D cuts at the diagonal direction in the $(x,z)$ plane (i.e.,
at about $45^\mathrm{o}$
as functions of time.
The right panel of Figure~\ref{hc} shows the evolution in the $(H,C)$ plane
of the mean values
compared to the (4-point) prediction for the fractional Brownian motion series  \citep{bash07,zunial08}.
Figure~\ref{hc} (top) shows that as turbulence develops
the permutation entropy $H$ becomes about $0.88$ 
whereas $C$ reaches values around $0.13$ and indicates
 a long-time trend of
 $H$ to increase and $C$ to decreases, as might be  expected.
 However,
when the fire hose wave activity appears around $t=0.14 t_{\text{exp}}$
 $H$ is temporarily 
reduced and $C$ enhanced; these changes are significant,
they are somewhat larger than the corresponding standard deviations.
The mean values of the permutation entropy $H$ and the statistical complexity $C$ 
are anti-correlated and (except for the initial phase when turbulence
develops) $H$ and $C$ follow a path compatible with
the prediction for the fractional Brownian motion \cite[cf.,][]{mamo13}
(Figure~\ref{hc}, right).
We also calculated 
the permutation entropy and statistical complexity for 1-D cuts
parallel and perpendicular to the background magnetic field;
calculations along these direction give similar results. However,
in the perpendicular direction the variation of $H$ and $C$ due
to the fire hose wave activity is rather weak. 

We performed an additional simulation of a decaying (only 2-D) turbulence 
with similar parameters but far from the unstable region 
using the expanding box model. In this simulation
no fire hose wave activity is observed and the mean values of $H$ and 
 $C$ (calculated on 1-D spatial cuts along x
of the $B_y$ component) have the expected behavior:
 once the turbulent cascade is well developed, $H$
slowly increases whereas $C$ decreases (not shown here).
We observe a similar evolution in a corresponding standard hybrid simulation.
This supports the idea that the presence of quasi-coherent fire hose wave
activity leads to a decrease of the permutation entropy and an increase
of the complexity.

\section{Discussion \& Conclusion}

In this paper we used the 3-D HEB simulation to investigate
properties of the turbulent cascade and its relationship
with fire hose instabilities. We initialized
the system with an isotropic spectrum of (relatively) large-scale Alfv\'enic
fluctuations having zero cross-helicity. Nonlinear coupling of these
modes leads to a turbulent cascade and energization of protons.
Such energization is not, however, strong enough to compete
with the anisotropic cooling due to the expansion and
protons develop an important parallel ($T_{\perp} < T_{\|}$) temperature anisotropy
similar to in situ observations from Helios spacecraft \citep{mattal07}; larger-amplitude turbulent
fluctuations are needed to counterbalance the solar wind expansion \cite[cf.,][]{montal18}. 
When the proton temperature anisotropy becomes strong enough, parallel and oblique fire hoses
are destabilized and generate quasi-coherent wave packets with an amplitude
much smaller than that of the background turbulence. 
The generated waves efficiently reduce the proton temperature anisotropy,
they are partly reabsorbed by protons through cyclotron resonances
and partly couple to and participate in the turbulent cascade. 
Turbulent fluctuations shape the plasma system, leading
to the formation of variable electromagnetic field as
well as variable proton density, temperature anisotropy,
agyrotropy, etc. The fire hose wave activity reduces the
proton temperature anisotropy while it further increases
the proton agyrotropy. Despite the inhomogeneity/nonuniformity and
agyrotropy of the plasma system, the linear prediction
based on the corresponding homogeneous and uniform gyrotropic approximation
is in a semi-quantitative agreement with the simulation results.

In the 2-D HEB simulation of \cite{hellal15} the expansion-driven fire hose fluctuations
are by construction injected to the turbulent cascade; they are forced to have
wave vectors perpendicular to the ambient magnetic field where the cascade proceeds.
In this case, the simulation results indicate that the growth rate of unstable modes must be strong
enough to compete with the cascade characterized by the nonlinear eddy turn-over time
$t_{nl}$. In the real space, the fire hose fluctuations were strongly localized
in the 2-D simulation between magnetic islands and tend to be aligned along the local most uniform direction.
In the 3-D simulation the fire hose fluctuations appear
to be outside the region dominated by the turbulent cascade; their 
wave vectors are quasi-parallel/oblique with
respect to the ambient magnetic field and lie outside of
the region in the wave vector space $(k_\|,k_\perp)$
dominated by the turbulent cascade. In this quasi-parallel/oblique
region the nonlinear eddy turn-over time ($t_{nl} \propto 1/\delta B$) is long
so that it does not present a large obstacle to the development of the instabilities.
In the real space, the fluctuations have quasi-parallel/oblique propagation and
are directed about along the ambient magnetic field (magnetic field-lines) that constitutes also roughly
the most uniform direction. We are inclined to speculate that
the presence of the turbulent cascade has a tendency to push the
unstable modes outside the region dominated by the cascade. 
More work is needed to test it.

The well-developed turbulence in the 3-D simulation before the onset of 
the fire hose instabilities exhibit anisotropic intermittency properties
that correspond to the anisotropic cascade.
The kurtosis depends strongly on the separation scale $l_\perp$ perpendicular to the
ambient magnetic field
\cite[similar results are obtained also for the standard 3-D hybrid simulation of][]{franal18a}.
The kurtosis saturates and decreases for scales  $l_\perp \lesssim d_i$ and is weaker
at the quasi-parallel direction $l_\| > l_\perp$. The reduction of the kurtosis is
likely related to the (resistive) dissipation, a relatively large resistivity is used
to avoid the accumulation of the magnetic energy on small scales. 
The weak fire hose wave activity affects the kurtosis  and
leads to its reduction at the quasi-parallel
directions; the generated waves have likely rather Gaussian character which leads
to a reduction of kurtosis; however, because of their weak amplitude, they
do not affect significantly the statistical properties of fluctuations at oblique
angles where the turbulent fluctuations dominate.

For the developed turbulence in the presented simulation before the onset of
the fire hose instabilities the Shannon entropy $H$ and the Jensen-Shannon complexity $C$
comparable to in situ observations in the
solar wind \citep{weckal15,weki19}, $H$ increases and $C$ increases with the time (corresponding
to the radial distance)
as expected (and observed in a simulation where the fire hose instabilities are not
generated) and they follow the prediction
for a fractional Brownian motion in the $(H,C)$ plane similarly to
the observational results of \cite{weki19}.
The fire hose wave activity brings new information to the system and causes a decreases of the entropy $H$ 
and to an increase of the complexity $C$ as may be expected. 
Interestingly, even when the fire hose waves are present, 
$H$ and $C$ keep following the prediction for a fractional Brownian motion.
However, the meaning of $H$ and $C$ (and their connection to the fractional Brownian motion) 
are yet to be understood.

This wave activity is easily identified as a narrow peak on top of
the background turbulence in 1-D power spectra
similar to observations but only at quasi-parallel/oblique
angles with respect to 
the ambient magnetic field. For more oblique angles,
the fire hose wave activity is not discernible because
of its low amplitude compared to the turbulent background.
We expect a similar behavior for in situ observations. In
this case it may be possible to detect the wave activity
based on the coherence property \citep{lional16}.
Furthermore, as the instability-driven wave activity modifies the statistical
properties of the turbulent fluctuations, one may ask if
these quantities may be used to discern generation
of a (quasi-coherent) wave activity (such as instability-driven waves)
within a turbulent system. It seems that the intermittency/kurtosis is not very useful in this
respect, as it is only reduced in the quasi-parallel direction.  
The  Shannon  entropy and
the Jensen-Shannon complexity are more promising tools;
in contrast to the kurtosis, they don't seem to be strongly dependent on the
angle the time series are measured with respect to the ambient magnetic field. 
However, it is unclear whether they can be used for real observations,
since the change of two quantities in the simulation is rather weak.

The present 3-D simulation has many limitations.
 We used parameters similar to the 2D case of \cite{hellal15}
with an expansion about ten times faster than in the solar wind. At the
same time, the amplitude of the turbulent fluctuations were larger so that 
the ratio between the expansion and turbulence characteristic time scales
is comparable to that observed in the solar wind.  
The fire hose driven magnetic fluctuations reach a relatively low amplitude
with respect to the ambient turbulence but they are efficient in reducing
the proton temperature anisotropy owing to the resonant character of their
interaction with protons. For a realistic expansion time we expect even a weaker level
of fluctuations \citep{mattal06,hell17}; the realistic level of turbulent
fluctuation is also lower so that we expect a similar relative amplitude
of fire hose and turbulent fluctuations in the solar wind.
A probably more important limitation of the present results
is the size of the simulation box.
Due to the limited resources we have a relatively small box size and we initially
inject the energy at relatively small scales compared to the solar wind conditions. 
Consequently, the level of intermittency
observed in the simulation is relatively weak. Its formation on small scales is also possibly
reduced by noise due to the limited number of particles per cell.

Despite these limitations, our numerical results show that 
fire hose instabilities coexist with plasma turbulence
and generate temperature-anisotropy reducing
modes that lie in the spectral space
 outside the region dominated by the turbulent cascade.
We expect a similar behavior for other
ion kinetic instabilities driven by particle temperature anisotropies
and/or differential streaming in the solar wind
\cite[as indicated by in situ observations of][]{kleinal18},
 except possibly for the
mirror instability that drives strongly oblique modes; 
2-D numerical simulations 
show that the mirror instability coexists
with plasma turbulence \citep{hellal17}. However, the present
simulation results show that these 2-D results are too limited
and need to be revisited.

\acknowledgments
PH acknowledges grant 18-08861S of the Czech Science Foundation.
LF was supported by the UK Science and Technology Facilities Council (STFC) grant ST/P000622/1.
This work was supported by the Programme National PNST of CNRS/INSU co-funded by CNES.
We thank T. S. Plachutta for useful comments and suggestions.

\bibliographystyle{apj2-eid}

\begin{thebibliography}{}
\expandafter\ifx\csname natexlab\endcsname\relax\def\natexlab#1{#1}\fi

\bibitem[{Alexandrova {et~al.}(2008)Alexandrova, Carbone, Veltri, \&
  Sorriso-Valvo}]{alexal08}
Alexandrova, O., Carbone, V., Veltri, P., \& Sorriso-Valvo, L. 2008, ApJ, 674,
  1153, doi:10.1086/524056

\bibitem[{Alexandrova {et~al.}(2013)Alexandrova, Chen, Sorriso-Valvo, Horbury,
  \& Bale}]{alexal13}
Alexandrova, O., Chen, C. H.~K., Sorriso-Valvo, L., Horbury, T.~S., \& Bale,
  S.~D. 2013, Space Sci. Rev., 178, 101, doi:10.1007/s11214-013-0004-8

\bibitem[{Arzamasskiy {et~al.}(2019)Arzamasskiy, Kunz, Chandran, \&
  Quataert}]{arzaal19}
Arzamasskiy, L., Kunz, M.~W., Chandran, B. D.~G., \& Quataert, E. 2019, ApJ,
  879, 53, doi:10.3847/1538-4357/ab20cc

\bibitem[{Bandt \& Pompe(2002)}]{bapo02}
Bandt, C., \& Pompe, P. 2002, PhRvL, 88, 174102

\bibitem[{Bandt \& Shiha(2007)}]{bash07}
Bandt, C., \& Shiha, F. 2007, J. Time Ser. Analysis, 28, 646,
  doi:10.1111/j.1467-9892.2007.00528.x

\bibitem[{Bruno \& Carbone(2013)}]{brca13}
Bruno, R., \& Carbone, V. 2013, LRSP, 10, 2, doi:10.12942/lrsp-2013-2

\bibitem[{Cranmer {et~al.}(2009)Cranmer, Matthaeus, Breech, \&
  Kasper}]{cranal09}
Cranmer, S.~R., Matthaeus, W.~H., Breech, B.~A., \& Kasper, J.~C. 2009, ApJ,
  702, 1604, doi:10.1088/0004-637X/702/2/1604

\bibitem[{{Del Sarto} \& Pegoraro(2018)}]{depe18}
{Del Sarto}, D., \& Pegoraro, F. 2018, MNRAS, 475, 181,
  doi:10.1093/mnras/stx3083

\bibitem[{Dong {et~al.}(2014)Dong, Verdini, \& Grappin}]{dongal14}
Dong, Y., Verdini, A., \& Grappin, R. 2014, ApJ, 793, 118,
  doi:10.1088/0004-637X/793/2/118

\bibitem[{Franci {et~al.}(2018{\natexlab{a}})Franci, Hellinger, Guarrasi, Chen,
  Papini, Verdini, Matteini, \& Landi}]{franal18b}
Franci, L., Hellinger, P., Guarrasi, M., {et~al.} 2018{\natexlab{a}}, J. Phys.:
  Conf. Ser., 1031, 012002, doi:10.1088/1742-6596/1031/1/012002

\bibitem[{Franci {et~al.}(2015)Franci, Landi, Matteini, Verdini, \&
  Hellinger}]{franal15b}
Franci, L., Landi, S., Matteini, L., Verdini, A., \& Hellinger, P. 2015, ApJ,
  812, 21, doi:10.1088/0004-637X/812/1/21

\bibitem[{Franci {et~al.}(2018{\natexlab{b}})Franci, Landi, Verdini, Matteini,
  \& Hellinger}]{franal18a}
Franci, L., Landi, S., Verdini, A., Matteini, L., \& Hellinger, P.
  2018{\natexlab{b}}, ApJ, 1, 26, doi:10.3847/1538-4357/aaa3e8

\bibitem[{Gary(1993)}]{gary93}
Gary, S.~P. 1993, Theory of Space Plasma Microinstabilities (New York:
  Cambridge Univ. Press)

\bibitem[{Gary {et~al.}(2001)Gary, Goldstein, \& Steinberg}]{garyal01}
Gary, S.~P., Goldstein, B.~E., \& Steinberg, J.~T. 2001, JGR, 106, 24955

\bibitem[{{Gary} {et~al.}(2016){Gary}, {Jian}, {Broiles}, {Stevens}, {Podesta},
  \& {Kasper}}]{garyal16}
{Gary}, S.~P., {Jian}, L.~K., {Broiles}, T.~W., {et~al.} 2016, JGR, 121, 30,
  doi:10.1002/2015JA021935

\bibitem[{Gary {et~al.}(1998)Gary, Li, O'Rourke, \& Winske}]{garyal98}
Gary, S.~P., Li, H., O'Rourke, S., \& Winske, D. 1998, JGR, 103, 14567

\bibitem[{Grappin {et~al.}(1993)Grappin, Velli, \& Mangeney}]{grapal93}
Grappin, R., Velli, M., \& Mangeney, A. 1993, PhRvL, 70, 2190

\bibitem[{Hellinger(2017)}]{hell17}
Hellinger, P. 2017, JPlPh, 83, 705830105, doi:10.1017/S0022377817000071

\bibitem[{Hellinger {et~al.}(2017)Hellinger, Landi, Matteini, Verdini, \&
  Franci}]{hellal17}
Hellinger, P., Landi, S., Matteini, L., Verdini, A., \& Franci, L. 2017, ApJ,
  838, 158, doi:10.3847/1538-4357/aa67e0

\bibitem[{Hellinger \& Matsumoto(2000)}]{hema00}
Hellinger, P., \& Matsumoto, H. 2000, JGR, 105, 10519

\bibitem[{Hellinger \& Matsumoto(2001)}]{hema01}
---. 2001, JGR, 106, 13215

\bibitem[{Hellinger {et~al.}(2015)Hellinger, Matteini, Landi, Verdini, Franci,
  \& {Tr\'avn\'{\i}\v{c}ek}}]{hellal15}
Hellinger, P., Matteini, L., Landi, S., {et~al.} 2015, ApJL, 811, L32,
  doi:10.1088/2041-8205/811/2/L32

\bibitem[{Hellinger \& {Tr\'avn\'{\i}\v{c}ek}(2005)}]{hetr05}
Hellinger, P., \& {Tr\'avn\'{\i}\v{c}ek}, P. 2005, JGR, 110, A04210,
  doi:10.1029/2004JA010687

\bibitem[{Hellinger \& {Tr\'avn\'{\i}\v{c}ek}(2008)}]{hetr08}
---. 2008, JGR, 113, A10109, doi:10.1029/2008JA013416

\bibitem[{Hellinger {et~al.}(2006)Hellinger, {Tr\'avn\'{\i}\v{c}ek}, Kasper, \&
  Lazarus}]{hellal06}
Hellinger, P., {Tr\'avn\'{\i}\v{c}ek}, P., Kasper, J.~C., \& Lazarus, A.~J.
  2006, GeoRL, 33, L09101, doi:10.1029/2006GL025925

\bibitem[{Hellinger {et~al.}(2013)Hellinger, {Tr\'avn\'{\i}\v{c}ek}, {\v
  S}tver\'ak, Matteini, \& Velli}]{hellal13}
Hellinger, P., {Tr\'avn\'{\i}\v{c}ek}, P.~M., {\v S}tver\'ak, {\v S}.,
  Matteini, L., \& Velli, M. 2013, JGR, 118, 1351, doi:10.1002/jgra.50107

\bibitem[{Jian {et~al.}(2009)Jian, Russell, Luhmann, Strangeway, Leisner, \&
  Galvin}]{jianal09}
Jian, L.~K., Russell, C.~T., Luhmann, J.~G., {et~al.} 2009, ApJL, 701, L105,
  doi:10.1088/0004-637X/701/2/L105

\bibitem[{{Klein} {et~al.}(2018){Klein}, {Alterman}, {Stevens}, {Vech}, \&
  {Kasper}}]{kleinal18}
{Klein}, K.~G., {Alterman}, B.~L., {Stevens}, M.~L., {Vech}, D., \& {Kasper},
  J.~C. 2018, PhRvL, 120, 205102, doi:10.1103/PhysRevLett.120.205102

\bibitem[{{Klein} \& {Howes}(2015)}]{klho15}
{Klein}, K.~G., \& {Howes}, G.~G. 2015, PhPl, 22, 032903, doi:10.1063/1.4914933

\bibitem[{Lamberti {et~al.}(2004)Lamberti, Martin, Plastino, \&
  Rosso}]{lambal04}
Lamberti, P.~W., Martin, M.~T., Plastino, A., \& Rosso, O.~A. 2004, Phys. A,
  334, 119

\bibitem[{Lion {et~al.}(2016)Lion, Alexandrova, \& Zaslavsky}]{lional16}
Lion, S., Alexandrova, O., \& Zaslavsky, A. 2016, ApJ, 824, 47,
  doi:10.3847/0004-637X/824/1/47

\bibitem[{{L{\'o}pez-Ruiz} {et~al.}(1995){L{\'o}pez-Ruiz}, Mancini, \&
  Calbet}]{lopeal95}
{L{\'o}pez-Ruiz}, R., Mancini, H.~L., \& Calbet, X. 1995, Phys. Lett. A, 209,
  321, doi:10.1016/0375-9601(95)00867-5

\bibitem[{MacBride {et~al.}(2008)MacBride, Smith, \& Forman}]{macbal08}
MacBride, B.~T., Smith, C.~W., \& Forman, M.~A. 2008, ApJ, 679, 1644,
  doi:10.1086/529575

\bibitem[{Maggs \& Morales(2013)}]{mamo13}
Maggs, J.~E., \& Morales, G.~J. 2013, Plasma Phys. Control. Fusion, 55, 085015,
  doi:10.1088/0741-3335/55/8/085015

\bibitem[{Markovskii \& Vasquez(2011)}]{mava11}
Markovskii, S.~A., \& Vasquez, B.~J. 2011, ApJ, 739, 22,
  doi:10.1088/0004-637X/739/1/22

\bibitem[{Maruca {et~al.}(2012)Maruca, Kasper, \& Gary}]{marual12}
Maruca, B.~A., Kasper, J.~C., \& Gary, S.~P. 2012, ApJ, 748, 137,
  doi:10.1088/0004-637X/748/2/137

\bibitem[{Matteini {et~al.}(2013)Matteini, Hellinger, Goldstein, Landi, Velli,
  \& Neugebauer}]{mattal13}
Matteini, L., Hellinger, P., Goldstein, B.~E., {et~al.} 2013, JGR, 118, 2771,
  doi:10.1002/jgra.50320

\bibitem[{Matteini {et~al.}(2012)Matteini, Hellinger, Landi,
  {Tr\'avn\'{\i}\v{c}ek}, \& Velli}]{mattal12}
Matteini, L., Hellinger, P., Landi, S., {Tr\'avn\'{\i}\v{c}ek}, P.~M., \&
  Velli, M. 2012, Space Sci. Rev., 172, 373, doi:10.1007/s11214-011-9774-z

\bibitem[{Matteini {et~al.}(2007)Matteini, Landi, Hellinger, Pantellini,
  Maksimovic, Velli, Goldstein, \& Marsch}]{mattal07}
Matteini, L., Landi, S., Hellinger, P., {et~al.} 2007, GeoRL, 34, L20105,
  doi:10.1029/2007GL030920

\bibitem[{Matteini {et~al.}(2006)Matteini, Landi, Hellinger, \&
  Velli}]{mattal06}
Matteini, L., Landi, S., Hellinger, P., \& Velli, M. 2006, JGR, 111, A10101,
  doi:10.1029/2006JA011667

\bibitem[{Matthaeus \& Velli(2011)}]{mave11}
Matthaeus, W.~H., \& Velli, M. 2011, Space Sci. Rev., 160, 145,
  doi:10.1007/s11214-011-9793-9

\bibitem[{Matthaeus {et~al.}(2015)Matthaeus, Wan, Servidio, Greco, Osman,
  Oughton, \& Dmitruk}]{mattal15d}
Matthaeus, W.~H., Wan, M., Servidio, S., {et~al.} 2015, Phil. Trans. R. Soc. A,
  373, 20140154, doi:10.1098/rsta.2014.0154

\bibitem[{Matthews(1994)}]{matt94}
Matthews, A. 1994, JCoPh, 112, 102

\bibitem[{{Mininni} \& {Pouquet}(2009)}]{mipo09}
{Mininni}, P.~D., \& {Pouquet}, A. 2009, PhRvE, 80, 025401,
  doi:10.1103/PhysRevE.80.025401

\bibitem[{Montagud-Camps {et~al.}(2018)Montagud-Camps, Grappin, \&
  Verdini}]{montal18}
Montagud-Camps, V., Grappin, R., \& Verdini, A. 2018, ApJ, 853, 153,
  doi:10.3847/1538-4357/aaa1ea

\bibitem[{Ofman(2019)}]{ofma19}
Ofman, L. 2019, Sol. Phys., 294, 51, doi:10.1007/s11207-019-1440-8

\bibitem[{Olivier {et~al.}(2019)Olivier, Engelbrecht, \& Strauss}]{olival19}
Olivier, C.~P., Engelbrecht, N.~E., \& Strauss, R.~D. 2019, JGR, 124, 4,
  doi:10.1029/2018JA026102

\bibitem[{Oughton {et~al.}(1994)Oughton, Priest, \& Matthaeus}]{oughal94}
Oughton, S., Priest, E.~R., \& Matthaeus, W.~H. 1994, J. Fluid Mech., 280, 95,
  doi:10.1017/S0022112094002867

\bibitem[{Papini {et~al.}(2019)Papini, Franci, Landi, Verdini, Matteini, \&
  Hellinger}]{papial19}
Papini, E., Franci, L., Landi, S., {et~al.} 2019, ApJ, 870, 52,
  doi:10.3847/1538-4357/aaf003

\bibitem[{{Parashar} {et~al.}(2018){Parashar}, {Matthaeus}, \&
  {Shay}}]{paraal18}
{Parashar}, T.~N., {Matthaeus}, W.~H., \& {Shay}, M.~A. 2018, ApJL, 864, L21,
  doi:10.3847/2041-8213/aadb8b

\bibitem[{Parashar {et~al.}(2009)Parashar, Shay, Cassak, \&
  Matthaeus}]{paraal09}
Parashar, T.~N., Shay, M.~A., Cassak, P.~A., \& Matthaeus, W.~H. 2009, PhPl,
  16, 032310, doi:10.1063/1.3094062

\bibitem[{Petrosyan {et~al.}(2010)Petrosyan, Balogh, Goldstein, {L\'eorat},
  Marsch, Petrovay, Roberts, {von Steiger}, \& Vial}]{petral10}
Petrosyan, A., Balogh, A., Goldstein, M.~L., {et~al.} 2010, Space Sci. Rev.,
  156, 135, doi:10.1007/s11214-010-9694-3

\bibitem[{Scudder \& Daughton(2008)}]{scda08}
Scudder, J., \& Daughton, W. 2008, JGR, 113, A06222, doi:10.1029/2008JA013035

\bibitem[{Servidio {et~al.}(2015)Servidio, Valentini, Perrone, Greco, Califano,
  Matthaeus, \& Veltri}]{serval15}
Servidio, S., Valentini, F., Perrone, D., {et~al.} 2015, JPlPh, 81, 325810107,
  doi:10.1017/S0022377814000841

\bibitem[{{Shebalin} {et~al.}(1983){Shebalin}, {Matthaeus}, \&
  {Montgomery}}]{shebal83}
{Shebalin}, J.~V., {Matthaeus}, W.~H., \& {Montgomery}, D. 1983, JPlPh, 29, 525

\bibitem[{Stansby {et~al.}(2019)Stansby, Perrone, Matteini, Horbury, \&
  Salem}]{stanal19}
Stansby, D., Perrone, D., Matteini, L., Horbury, T.~S., \& Salem, C.~S. 2019,
  A\&A, 623, L2, doi:10.1051/0004-6361/201834900

\bibitem[{{\v Stver\'ak} {et~al.}(2015){\v Stver\'ak}, {Tr\'avn\'{\i}\v{c}ek},
  \& Hellinger}]{stveal15}
{\v Stver\'ak}, {\v S}., {Tr\'avn\'{\i}\v{c}ek}, P.~M., \& Hellinger, P. 2015,
  JGR, 120, 8177, doi:10.1002/2015JA021368

\bibitem[{Valentini {et~al.}(2014)Valentini, Servidio, Perrone, Califano,
  Matthaeus, \& Veltri}]{valeal14}
Valentini, F., Servidio, S., Perrone, D., {et~al.} 2014, PhPl, 21, 082307,
  doi:10.1063/1.4893301

\bibitem[{{Verdini} \& {Grappin}(2015)}]{vegr15}
{Verdini}, A., \& {Grappin}, R. 2015, ApJL, 808, L34,
  doi:10.1088/2041-8205/808/2/L34

\bibitem[{Verdini {et~al.}(2015)Verdini, Grappin, Hellinger, Landi, \&
  {M\"uller}}]{verdal15}
Verdini, A., Grappin, R., Hellinger, P., Landi, S., \& {M\"uller}, W.~C. 2015,
  ApJ, 804, 119, doi:10.1088/0004-637X/804/2/119

\bibitem[{Weck {et~al.}(2015)Weck, Schaffner, Brown, \& Wicks}]{weckal15}
Weck, P.~J., Schaffner, D.~A., Brown, M.~R., \& Wicks, R.~T. 2015, PhRvE, 91,
  23101, doi:10.1103/PhysRevE.91.023101

\bibitem[{Weygand \& Kivelson(2019)}]{weki19}
Weygand, J.~M., \& Kivelson, M.~G. 2019, ApJ, 872, 59,
  doi:10.3847/1538-4357/aafda4

\bibitem[{{Wicks} {et~al.}(2016){Wicks}, {Alexander}, {Stevens}, {Wilson},
  {Moya}, {Vi{\~n}as}, {Jian}, {Roberts}, {O'Modhrain}, {Gilbert}, \&
  {Zurbuchen}}]{wickal16}
{Wicks}, R.~T., {Alexander}, R.~L., {Stevens}, M., {et~al.} 2016, ApJ, 819, 6,
  doi:10.3847/0004-637X/819/1/6

\bibitem[{Wu {et~al.}(2013)Wu, Wan, Matthaeus, Shay, \& Swisdak}]{wual13}
Wu, P., Wan, M., Matthaeus, W.~H., Shay, M.~A., \& Swisdak, M. 2013, PhRvL,
  111, 121105, doi:10.1103/PhysRevLett.111.121105

\bibitem[{Zunino {et~al.}(2008)Zunino, {P\'erez}, {Mart\'\i n}, Garavaglia,
  Plastino, \& Rosso}]{zunial08}
Zunino, L., {P\'erez}, D.~G., {Mart\'\i n}, M.~T., {et~al.} 2008, Phys. Lett.
  A, 372, 4768, doi:10.1016/j.physleta.2008.05.026

\end{thebibliography}

\end{document}